\documentclass[jou,apacite]{apa6}
\usepackage{times}
\usepackage{graphics,epsf}
\usepackage{amsmath}                
\usepackage{amsfonts}               
\usepackage{amssymb}                
\usepackage{epsfig}                 
\usepackage{graphicx}
\usepackage{url}
\usepackage{tabularx}
\usepackage{booktabs}
\usepackage{natbib}
\usepackage{rotating,multicol,adjustbox}

\title{AstroDance: Engaging Deaf and Hard-of-Hearing Students in Astrophysics via Multimedia Performances}
\shorttitle{APA style}

\sixauthors{J. Nordhaus, Ph.D.}{M. Campanelli, Ph.D.}{J. Bochner, Ph.D.}{T. Warfield, MFA}{H.-P. Bischof, Ph.D.}{J. Noel-Storr, Ph.D.}
\sixaffiliations{National Technical Institute for the Deaf\\
and\\
Center for Computational Relativity and Gravitation\\
Rochester Institute of Technology, Rochester, NY, USA}
{Center for Computational Relativity and Gravitation\\
and\\
Department of Mathematics\\
Rochester Institute of Technology, Rochester, NY, USA}
{Department of Cultural and Creative Studies\\
National Technical Institute for the Deaf\\
Rochester Institute of Technology, Rochester, NY, USA}
{Department of Performing Arts\\
National Technical Institute for the Deaf\\
Rochester Institute of Technology, Rochester, NY, USA}
{Center for Computational Relativity and Gravitation\\
and\\
Department of Computer Science\\
Rochester Institute of Technology, Rochester, NY, USA}
{Undergraduate School of Science and Engineering\\
and\\
Kapteyn Astronomical Institute\\
University of Groningen, Groningen, Netherlands
}

\abstract{The dynamics of gravitating astrophysical systems such as black holes and neutron stars are fascinatingly complex, offer some of nature's most spectacular phenomena, and capture the public's imagination in ways that few subjects can.  Here, we describe {\it AstroDance}, a multi-media project to engage deaf and hard-of-hearing (DHH) students in astronomy and gravitational physics.  {\it AstroDance} incorporates multiple means of representation of scientific concepts and was performed primarily for secondary and post-secondary audiences at $\sim$20 venues in the northeastern US prior to the historic first detection of gravitational waves.  As part of the {\it AstroDance} project, we surveyed $\sim$1000 audience members roughly split evenly between hearing and DHH audience members.  While both groups reported statistically equivalent high-rates of enjoyment of the performance, the DHH group reported an increase in how much they learned about science at a statistically significant rate compared to the hearing audience.  Our findings suggest that multi-sensory approaches benefit both hearing and deaf audiences and enable accessible participation for broader groups.}

\rightheader{AstroDance: Engaging Deaf and Hard-of-Hearing Students in Astrophysics}
\leftheader{Nordhaus}

\begin{document}
\maketitle
                        
\section{Introduction}
\label{sec:intro}

Deaf and hard-of-hearing (DHH) students are traditionally delayed in content-STEM learning areas compared to their hearing peers \citep{Marshark:2008ua}.  As such, DHH college student participation rates in STEM (Science, Technology, Engineering and Mathematics) fields lag as DHH students obtain STEM bachelor degrees at much lower rates than their hearing peers (15\% DHH vs. 25\% hearing; \citealt{Walter:2010xb}).  This in turn, leads to under-representation in the general STEM workforce, a group that benefits significantly from participation of diverse groups \citep{PCAST:2012ab}.

Multimedia scientific interactions at pre-collegiate ages could be one method to encourage longer-term interest in science \citep{DeLeoWinkler2019,deleowinkler:2019}.  Cognitive theories, such as dual-coding, suggest that learning is enhanced when acoustic, visual and linguistic signals are integrated as coding occurs on separate tracks \citep{Paivio}.  Multi-modal presentations provide additional sources of information, which may enhance memory and learning for audience members \citep{Paivio1971}.  Such multimedia scientific experiences should benefit both hearing and deaf participants because they can make use of auditory and/or visual information.

Communicating accurate scientific knowledge in an understandable way to secondary and post-secondary school students is challenging \citep{oritzgil:2011}.  Currently, there is limited science education research on deaf and hard-of-hearing students \citep{Moores2001,Mangrubang2004,Kurz2015,Trussell2018}.  Here, we describe {\it AstroDance}, an original dance and theatrical multi-modal performance designed to communicate gravitational-wave astronomy to deaf and hard-of-hearing audiences.  We first discuss how the historic discovery of gravitational-waves is opening a new window into the universe.  Following that, we discuss the development of {\it AstroDance}.  As part of the {\it AstroDance} project, data was collected from $\sim$1000 audience members equally split between DHH and hearing populations.  Our initial hypothesis was that an inclusive, multi-media presentation would produce similar learning experiences independent of hearing status.  We discuss our analysis of the audience data including quantitative and qualitative data before concluding.

\subsection{Why Gravitational Astronomy is Important}
\label{sec:gravwaves}

A hundred years ago, Einstein's theory of General Relativity introduced 
an entirely new way of regarding space and time.
Rather than being entirely separate, they became different facets of a single concept commonly referred to as ``spacetime".  General relativity is a geometric theory of gravitation in which the curvature of spacetime is directly related to the presence of matter and radiation.  Among its many astonishing 
predictions, general relativity postulated the existence of gravitational waves and   
black holes. 

On September 14th 2015 arguably the most important physics discovery of the last half century 
was uncovered when scientists working at the 
Laser Interferometer Gravitational-Wave Observatory (LIGO)
~\citep{TheLIGOScientific:2014jea,Dwyer:2015fua} 
directly detected gravitational waves from the collision of two black holes
~\citep{Abbott:2016blz, TheLIGOScientific:2016wfe,
  TheLIGOScientific:2016src, Abbott:2016apu, Abbott:2016nmj,
TheLIGOScientific:2016pea}.  Gravitational waves are ``ripples" in spacetime
produced when massive objects like black holes collide in distant regions of the Universe. As LIGO continues uncovering gravitational wave signals, we have, and continue to, uncover the unknown.

The gravitational waves that are observable by LIGO are caused by some 
of the most energetic events in the Universe -- colliding black holes, merging neutron stars, exploding stars, and even the birth of the Universe itself. Detecting and analyzing the information carried by gravitational waves allows us to observe the Universe in a way never before possible.

Historically, scientists have relied primarily on observations with electromagnetic radiation (visible light, x-rays, radio waves, microwaves, etc.) to learn about and understand objects and phenomena in the Universe.  

Gravitational waves however, are a completely different phenomenon than electromagnetic waves, carrying complementary information about cosmic objects and the events that generated them.  Gravitational waves are also virtually unimpeded since they interact very weakly 
with matter (unlike electromagnetic radiation) as they traverse intergalactic space, 
giving us a clear view of the gravitational-wave Universe. 
Colliding black holes, for example, emit little or no electromagnetic 
radiation, but the gravitational waves they emit will enable them to be seen from areas of the universe that were previously impossible to probe with traditional telescopes. 

With this completely new way of examining 
astrophysical objects and phenomena, gravitational waves have truly opened a new window 
on the Universe, providing astronomers and other scientists with their first glimpses of 
previously unseen and invisible wonders, and greatly adding to our understanding 
of the nature of space and time itself. The significance of this discovery is therefore 
of monumental importance, one that will requires the rewriting of science 
textbooks.

The announcement of the first detection of gravitational waves in 2016 became a world-wide sensation.  
For a brief moment, the physics of black holes and gravitational 
waves outshone all other news, generating a wave of positive coverage exciting the public consciousness.
The worldwide response to the announcement was not restricted to the mainstream media. 
The sheer breadth and depth of interest it generated was a testimony to the importance of the result. 
Newspaper and television news coverage of the gravitational wave detection included front-page articles 
in the New York Times, and coverage on CNN and the BBC. According to the Newseum, 
a total of 961 newspaper front pages from February 12 (the day after the announcement) featured the discovery, which included 
the ``Discovery of Gravitational Waves" on their list of dates in 2016 deemed to be of historical significance. 
In just a few days, there were more than 70 million tweets about the subject, including one from President Obama who tweeted @POTUS: ``Einstein was right! Congrats to @NSF and @LIGO on detecting gravitational waves 
-- a huge breakthrough in how we understand the universe".
These are just a few mentions of the media and public response that followed this historic discovery~\footnote{For more see the American Physical Society News Back Page \url{https://www.aps.org/publications/apsnews/201608/backpage.cfm}}.

While the initial excitement was widespread, many of the experiences described above were limited in scope (newspaper articles, tweets, or short TV clips) and often focused on LIGO itself or topics such as ``Einstein was right".  As an example, many members of the public did not realize that GW150914 was the first binary-black hole system ever discovered.  Furthermore, the concept of vibrations in spacetime was confusing, with many not fully comprehending that as the waves passed the Earth, the physical separation between the LIGO mirrors changes.  In essence, the distance between the mirrors compresses and expands as the gravitational waves pass the Earth.  Via AstroDance, we explored novel and lasting ways to deepen the public understanding of the astrophysical sources that produce gravitational waves and what it means for our understanding of the universe.

\subsection{Communicating Gravitational Wave Science to Deaf and Hard-of-Hearing Audiences}
\label{sec:comm}
In anticipation of the direct detection of gravitational waves and the plethora of future detections, we developed {\it AstroDance} -- an original dance and theatrical performance designed to communicate cutting-edge scientific research on gravitational astrophysics to diverse audiences such as deaf and hard-of-hearing students.  {\it AstroDance} was funded via a ``Communicating Research to Public Audiences" (CRPA) Division of Research Learning grant from the National Science Foundation.  The design of {\it AstroDance} leveraged the experience and talents of a diverse group of researchers, scientists, artists and educators at the Rochester Institute of Technology (RIT) to create an exciting, cutting-edge, multi-sensory representation of the scientific ideas behind gravitational waves and LIGO.  The performance was specifically designed to stimulate learning for broad audiences consisting of deaf and hard-of-hearing children, adults, and members of the general public.  During the performances, all audience members were given access to the most modern concepts of gravitational physics so as to enhance their interest in STEM fields.

The project team consisted of a unique collaboration among researchers at RIT's Center for Computational Relativity and Gravitation\footnote{http://ccrg.rit.edu} and theater artists, educators and scientists at RIT's National Technical Institute for the Deaf\footnote{http://www.ntid.rit.edu}. The general aim was to create an inclusive performance that would improve the public understanding of what happens in extreme astrophysical conditions of gravity and matter. Communication occurred through the use of advanced computation and visualization, dance and theater.  RIT's gravitational astrophysics group has made outstanding contributions to the success of numerical modeling of the GWs produced by the collisions of binary black holes and neutron stars. The group is also currently working to study electromagnetic emission from these sources. This modeling effort was already crucial for determining the nature of the GW150914 system and many of the subsequent gravitational-wave detections.  Utilizing visual imagery obtained from the RIT group's modeling and simulation of GW sources was a cornerstone of the project.  Such scientific information is cutting edge, timely and accurate, thereby providing the audience with pertinent scientific information while engaging them in an enjoyable and interesting learning experience.  

Another key component of the project is that for deaf and hard-of-hearing individuals who rely on American Sign Language as their primary means of communication, there is a pressing need for accessible, high-quality-educational physics content.  Automatic caption systems such as those employed by Google on Youtube, render most online videos unintelligible to deaf audiences. Even with proper captioning, the comprehension of online physics content can be very challenging for deaf individuals, many of whom have limited English language proficiency and rely heavily on visual and non-verbal sources of information for comprehension.  Our hypothesis was that designing a visual multi-modal learning experience that could engage deaf audiences would not only provide direct benefit but might also provide benefit to hearing members of the audience.

\begin{figure}[h!]
\begin{center}
\includegraphics[width=8cm,angle=0,clip=true]{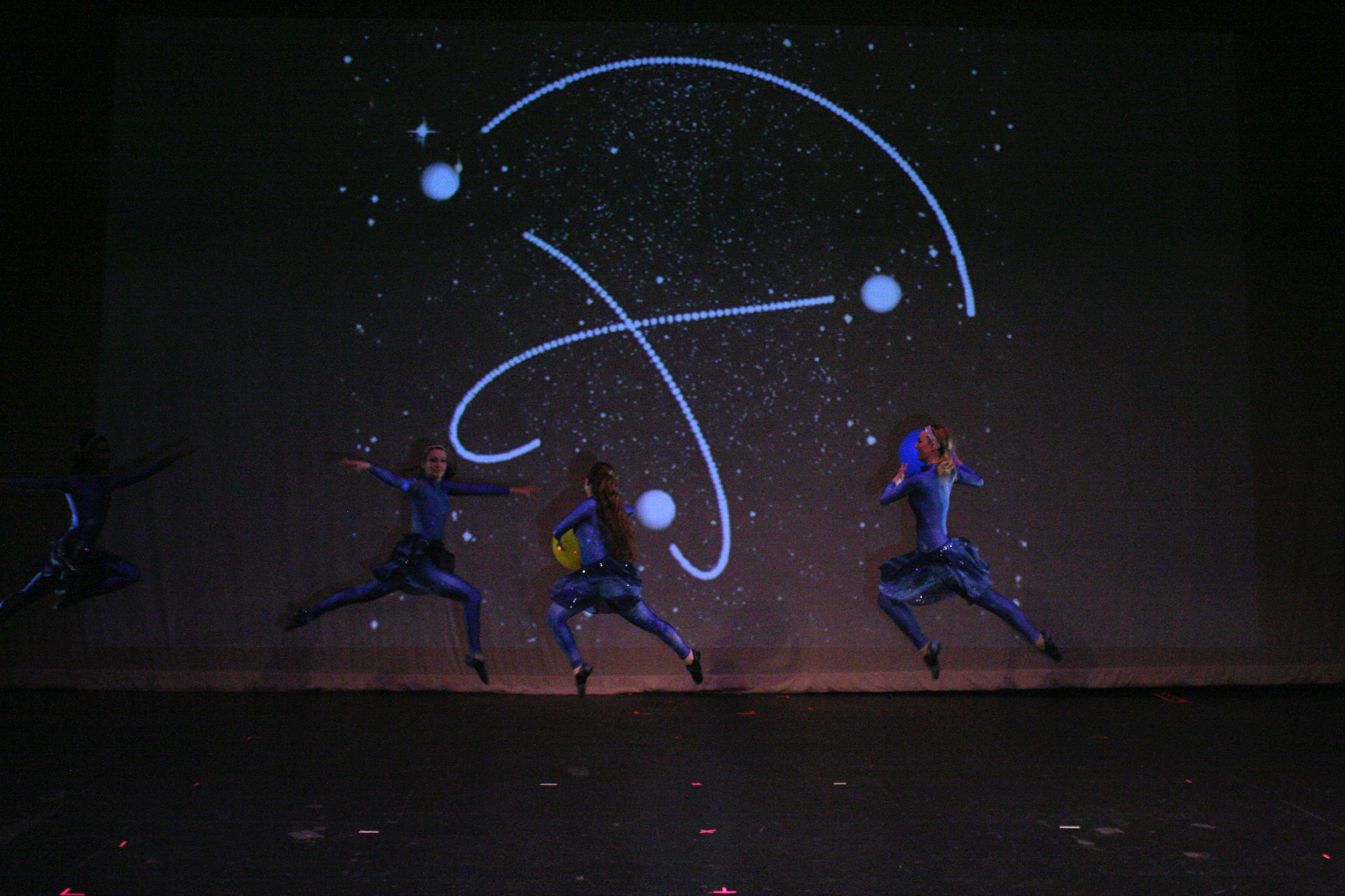}
\caption{Computer simulations of a tertiary black hole system, and a corresponding interpretative dance, showcase the complex dynamics of gravitating objects.
\label{fig:astrodance1}}
\end{center}
\end{figure}

\section{AstroDance Performances}
\label{sec:performances}

Our team of astrophysicists, science educators, dancers, computer scientists and choreographers at RIT worked to craft a visually appealing and accessible production from which diverse audiences could benefit.  The finished production, {\it AstroDance}, consisted of an overarching narrative that told the story of the hunt for gravitational waves while simultaneously communicating individual vignettes that highlighted the important aspects of gravitational science.  For each scientific concept, a short story was first narrated on-stage in both spoken English and American Sign Language.  A subsequent interpretative dance accompanied by music and accurate scientific visualizations (projected in the background) followed.  Each component of the vignette was designed to be complementary and reinforce the scientific concept.  For example, Figure~\ref{fig:astrodance1} shows {\it AstroDance} performers interpreting the movement of a black hole triple system as the black holes orbit and eventually coalesce producing gravitational waves.  Many of the scientific visualizations were referenced from scientific work done by members of the Center for Computational Relativity and Gravitation at RIT \citep{Nordhaus:2010ab,Nordhaus:2012aa,Nordhaus:2010aa,Campanelli:2006aa}.  

\begin{figure}[h!]
\begin{center}
\includegraphics[width=8cm,angle=0,clip=true]{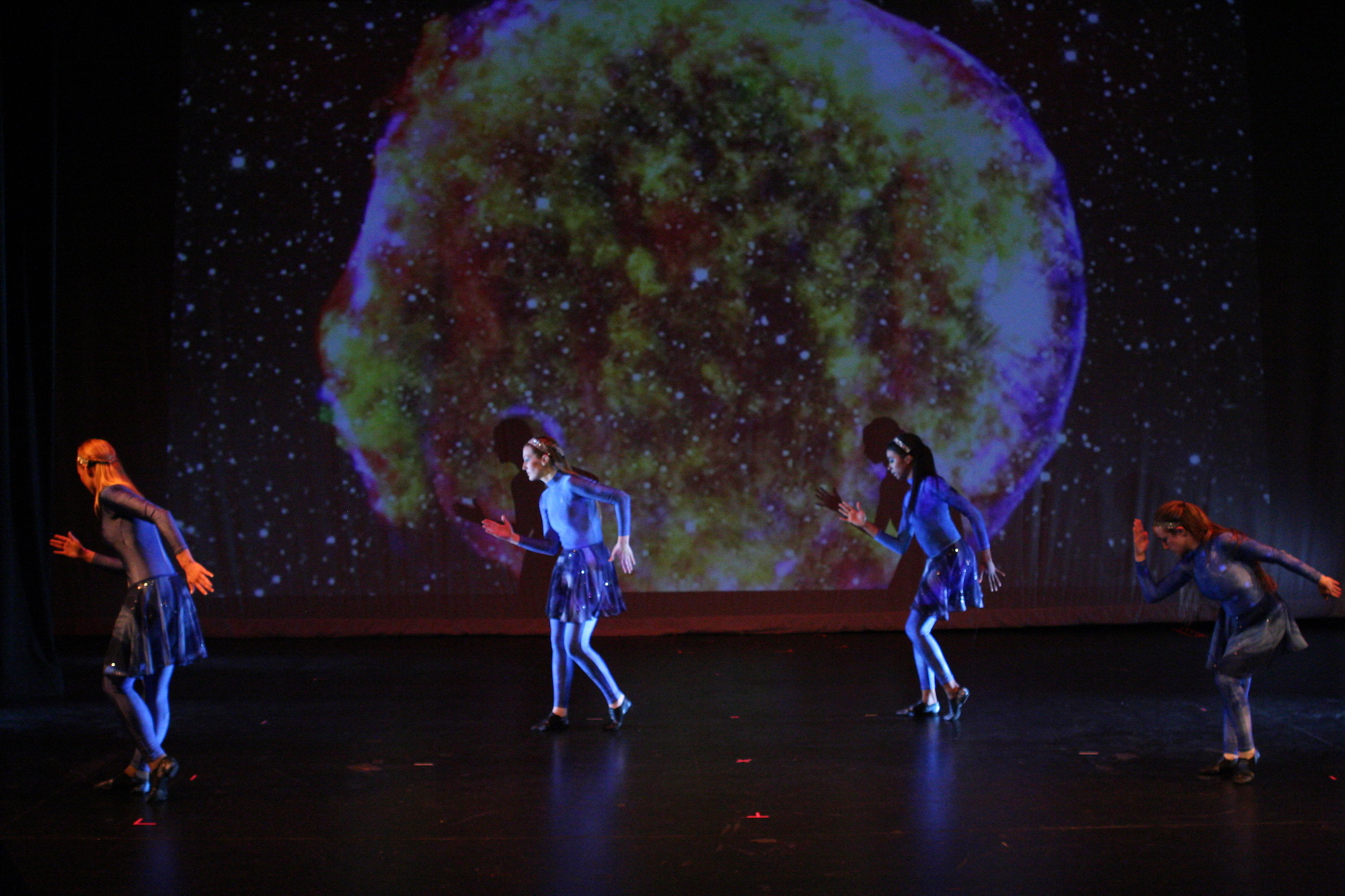}
\caption{{\it AstroDance} performers highlight the physics that occurs when a massive star collapses, and subsequently explodes as a supernova.
\label{fig:astrodance2}}
\end{center}
\end{figure}

\begin{figure}[h!]
\begin{center}
\includegraphics[width=8cm,angle=0,clip=true]{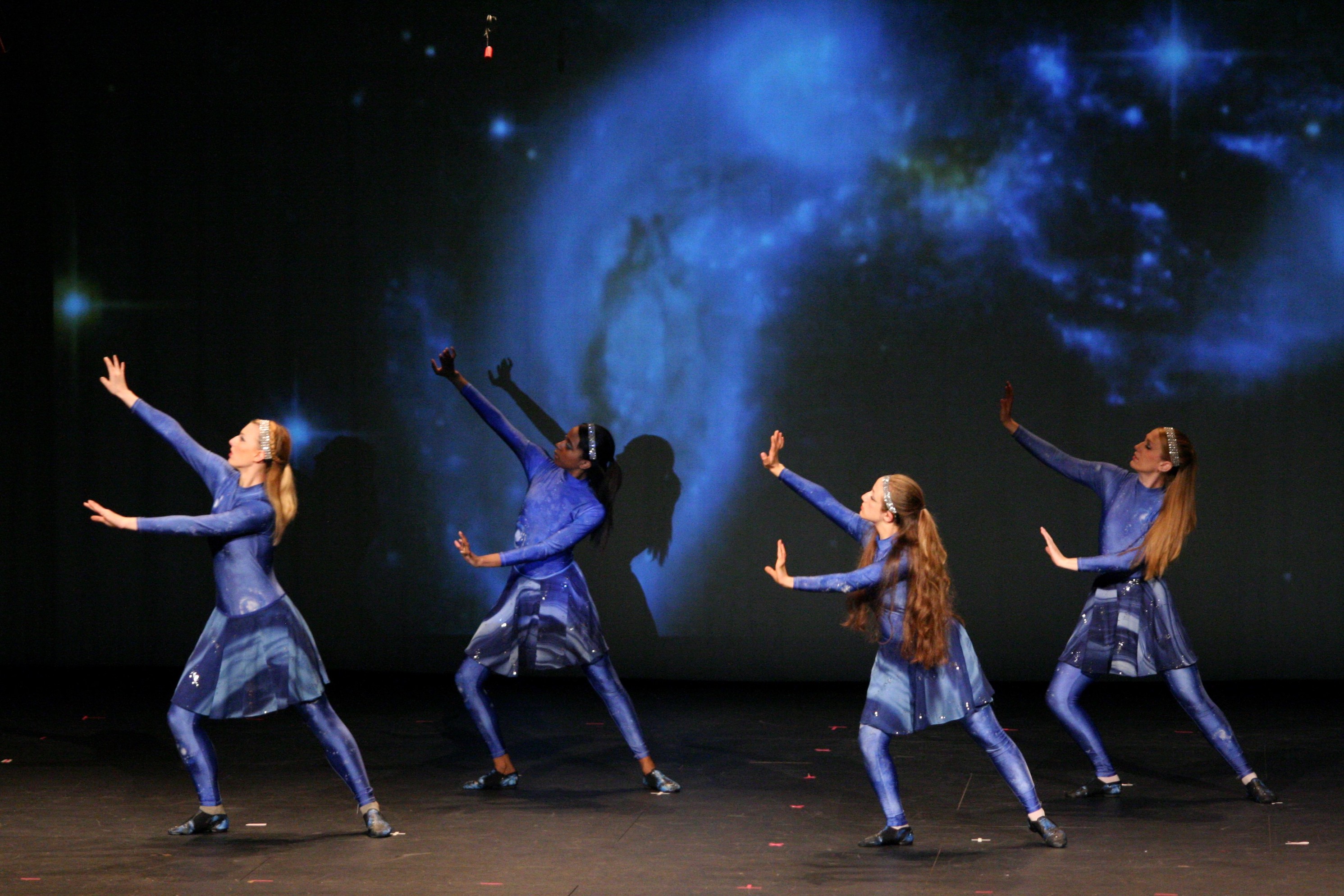}
\caption{The violent collisions between galaxies is depicted above.
\label{fig:astrodance3}}
\end{center}
\end{figure}

Figure~\ref{fig:astrodance2} depicts an interpretative dance of the implosion and subsequent explosion of a massive star.  Such core-collapse supernovae light up their host galaxies, produce most elements in nature and leave behind exotic remnants such as neutron stars and black holes.  Core-collapse events are an important future gravitational-wave source as LIGO may detect the gravitational waves emanating from a supernova in our galaxy.  The image in the background of Figure~\ref{fig:astrodance2} is of SN 1572, a Galactic supernova that occurred in November 1572 and was visible with the naked eye.  Figure~\ref{fig:astrodance3} shows an image from the interpretation of two colliding galaxies.  Such galactic mergers generate gravitational waves via the merger of their massive central black holes.

{\it AstroDance} premiered at the Little Theatre in Rochester NY on Sept. 22, 2012 as part of the Fringe Festival prior to the historic detection of gravitational waves.  A subsequent year-long tour was organized with {\it AstroDance} occurring in $\sim$20 venues in the Northeastern United States (see Table~\ref{table:performances}).  Note, that all AstroDance performances occurred at least three years prior to the announcement of the direct detection of gravitational waves.  AstroDance was designed to increase public awareness in anticipation of the historic detection with the caveat that one could not know when it would occur. At the end of each performance, a survey was administered to audience members which resulted in the extensive data presented in the following sections.

\section{Audience Survey and Feedback}
\label{sec:survey}

The primary metrics used to evaluate the outcomes of the project included the number people attending each performance, the number of visits to the project website and a survey administered to audience members at the completion of each performance for which Institutional Review Board approval was obtained.  An evaluation of the project website also was planned, but not completed due to unforeseen delays in the development of the website.  A list of all {\it AstroDance} performances with venue and date information is presented in Table~\ref{table:performances}.  {\it AstroDance} was performed at a range of secondary-school, collegiate and community venues throughout the northeastern United States.

Audience members were asked to complete a brief anonymous questionnaire at the conclusion of each performance.  Data collected included demographic information such as the audience member's age, gender, race/ethnicity, and hearing status.  The questionnaire content included three five-point Likert scales.  The scales asked audience members to rate the degree to which they: 1) enjoyed the performance; 2) learned about science; and 3) participate in science activities.  Note that since AstroDance was performed consistently at all venues, it is not possible to isolate the individual effects of dance, signed narration, or scientific visualizations.  Rather, the data collected applies to the aggregate effects of AstroDance in its entirety.  Finally, the survey asked audience members three open-ended questions.  One question asked respondents to describe the performance, another question asked what they learned from the performance, and the third question asked for other comments.

\subsection{Survey Demographics}
\label{sec:survey_demographics}

\begin{figure}[h!]
\begin{center}
\includegraphics[width=8cm,angle=0,clip=true]{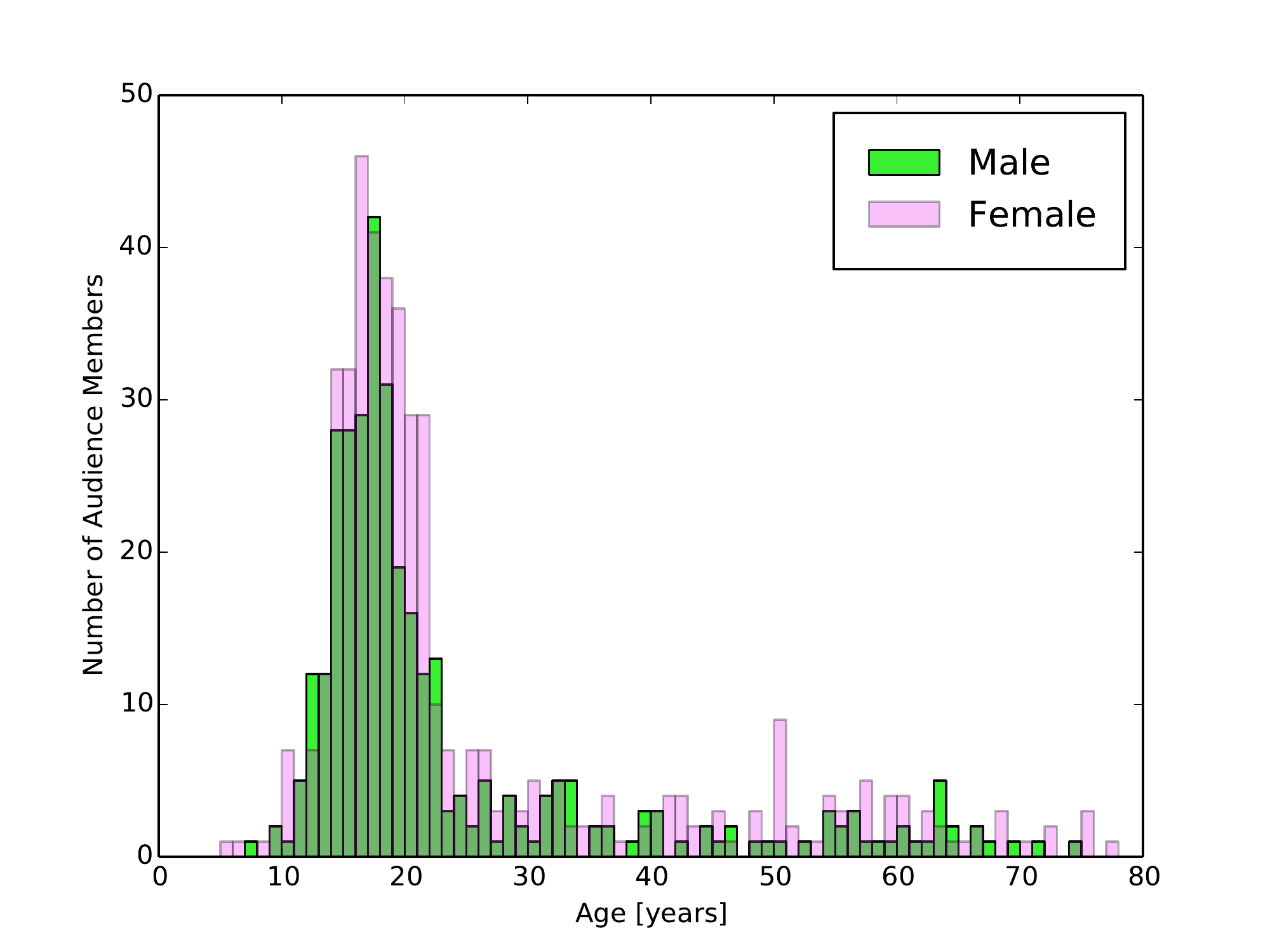}
\caption{Number of male and female audience members surveyed as a function of age for AstroDance performances.
\label{fig:age_histogram}}
\end{center}
\end{figure}

\begin{figure}[h!]
\begin{center}
\includegraphics[width=8cm,angle=0,clip=true]{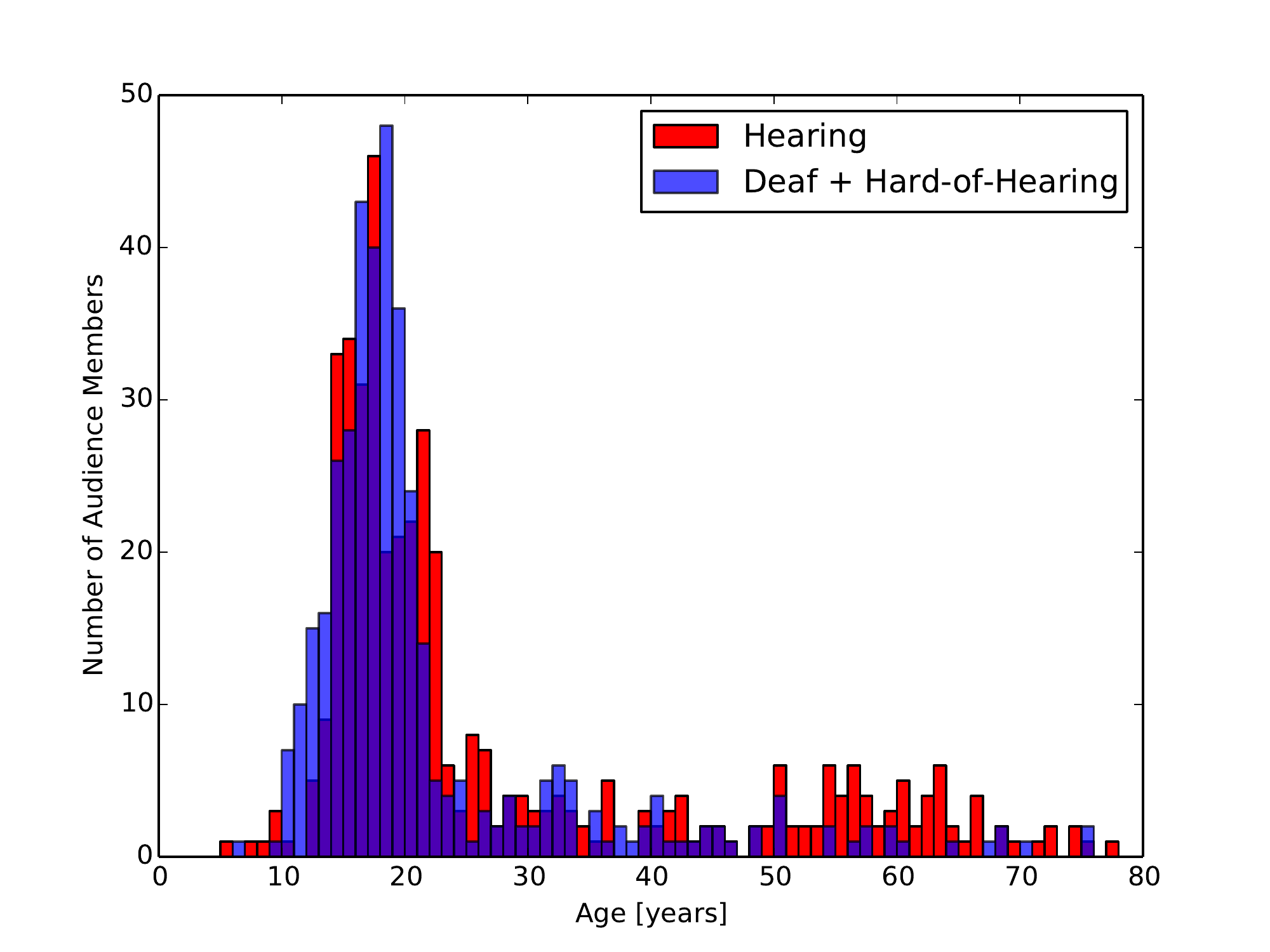}
\caption{Number of hearing and deaf/hard-of-hearing audience members surveyed as a function of age for AstroDance performances.
\label{fig:hearing_deaf_histogram}}
\end{center}
\end{figure}

We received 971 survey responses from audience members.  The distribution of gender for the audience as a function of age is presented as histograms in Figure~\ref{fig:age_histogram}.  Binning by age occurs in one-year increments with male audience members shown in green and female audience members in pink.  The bulk of the age distribution occurs between 13 and 21 years of age with the tail of the distribution primarily representing the age of their teachers.  The audience is roughly evenly divided between male and female audience members.  Figure~\ref{fig:hearing_deaf_histogram} shows the hearing status of the surveyed audience binned as in Figure~\ref{fig:age_histogram} with red representing hearing audience members and blue representing deaf and hard-of-hearing audience members.  The audience is roughly split evenly between hearing and deaf/hard-of-hearing participants.

Out of the 971 survey responses, 866 (89\%) self-reported demographic information on ethnicity.  Figure~\ref{fig:demographics} displays the ethnic breakdown with 44\% Caucasian, 26\% Hispanic, 16\% African American, 8\% Asian/Pacific Islander and 2\% Native Indian/Alaskan Native.

\subsection{Survey Results}
\label{sec:survey_results}

\begin{figure}[h!]
\begin{center}
\includegraphics[width=8cm,angle=0,clip=true]{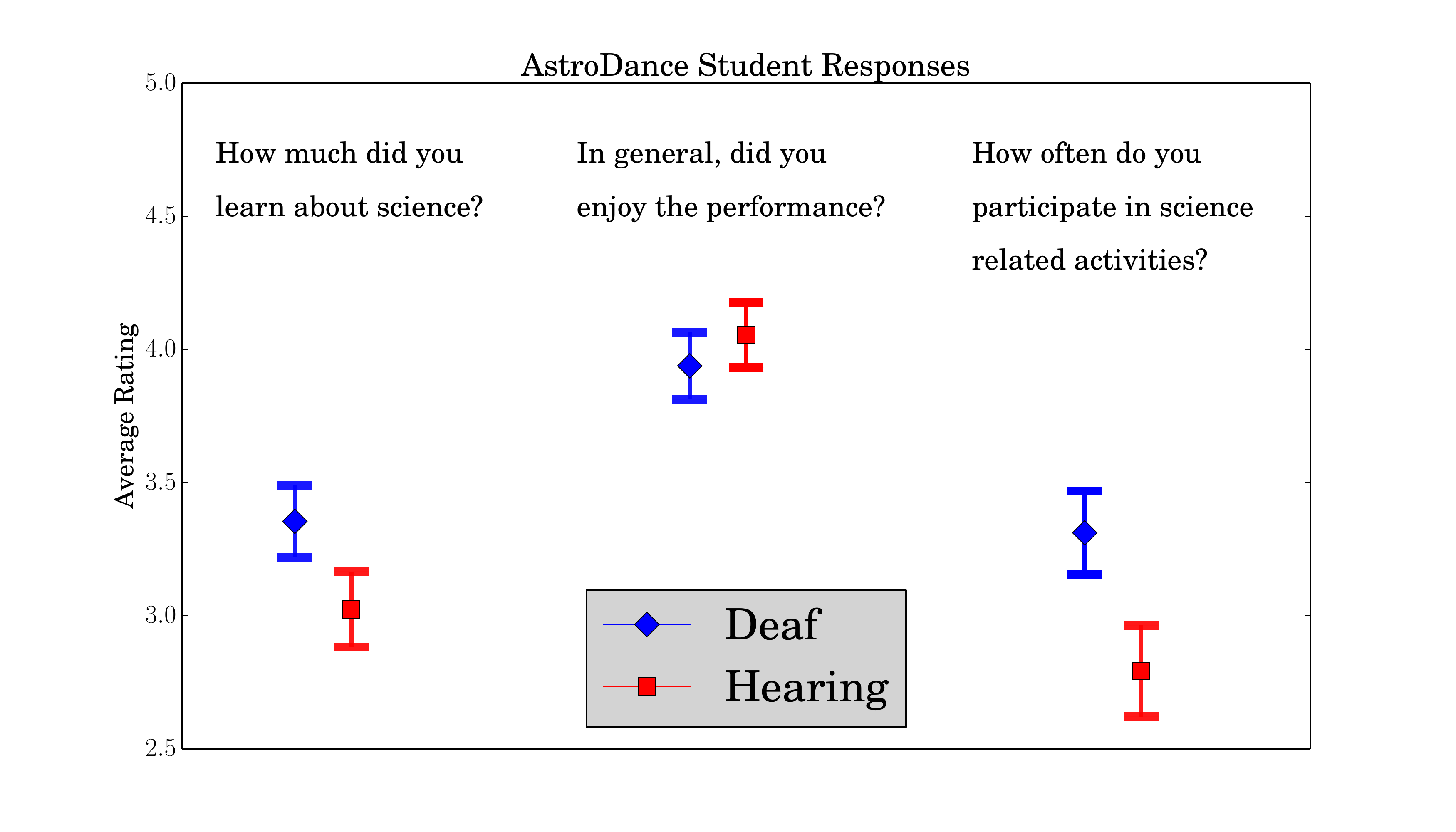}
\caption{DHH and hearing audience average responses to three Liekert-scale questions with 95\% confidence intervals and a total sample size of $N=563$.
\label{fig:response_averages}}
\end{center}
\end{figure}

To analyze the numerical scores for the three Liekert-scale questions, we decompose the audience into hearing and DHH student populations by imposing an age restriction of 22 years.  While our surveyed student audiences compose a blend of secondary-school and college students, this is a reasonable restriction given that part b of the Individuals with Disabilities Education Act (IDEA) provides special education services from ages 3 to 22.  Thus, even though we did not ask our audience to report student status below age 22, we would expect the vast majority of our responses to be from students.  Results are presented with error bars corresponding to 95\% confidence intervals in Figure~\ref{fig:response_averages}.  While both the hearing and DHH student groups reported that they learned from and enjoyed the performance, the deaf students reported a statistically significant higher response to ``How much did you learn about science?". To quantify the statistical significance of the difference in the means, we performed a Z-test and report a p-value of $0.001$.  This was the case despite the fact that the DHH student group also reported that they participated in science-related activities more often compared to the hearing student group (p-value of $0.00001$).  Both groups reported that they enjoyed the performances and were statistically indistinguishable (p-value of $0.19$).

\begin{table*}[t]
\begin{tabularx}{\textwidth}{p{8.5cm}p{4.5cm}lp{2.5cm}c}

\toprule
Performance Venue & City & Date \tabularnewline
\midrule
Little Theatre$^*$                                         & Rochester, NY       & 9/22/12 \tabularnewline
Rochester School for the Deaf$^\dagger$                    & Rochester, NY       & 9/28/12 \tabularnewline
NTID Dance Lab$^\ddagger$                                  & Rochester, NY       & 11/6/12 \tabularnewline
Western Pennsylvania School for the Deaf$^\dagger$         & Pittsburgh, PA      & 11/19/12 \tabularnewline
Scranton School for the Deaf and Hard-of-Hearing$^\dagger$ & Clarks Summit, PA   & 11/20/12 \tabularnewline
Scranton High School$^\dagger$                             & Scranton, PA        & 11/20/12 \tabularnewline
Gallaudet University$^\ddagger$                            & Washington, D.C.    & 1/26/13 \tabularnewline
University of Maryland Baltimore County$^\ddagger$         & Baltimore, MD       & 2/25/13 \tabularnewline
University of Rochester$^\ddagger$                         & Rochester, NY       & 3/29/13 \tabularnewline
Rochester Museum and Science Center$^*$                    & Rochester, NY       & 4/14/13 \tabularnewline
IS 47: The ASL \& English Secondary School$^\dagger$        & New York, NY        & 4/26/13 \tabularnewline
CUNY Graduate Center$^\ddagger$                            & New York, NY        & 4/26/13 \tabularnewline
Imagine RIT$^*$                                            & Rochester, NY       & 5/4/13 \tabularnewline
New York School for the Deaf - Fanwood$^\dagger$           & White Plains, NY    & 5/28/13 \tabularnewline
Lexington School for the Deaf$^\dagger$                    & Jackson Heights, NY & 5/29/13 \tabularnewline
Delaware School for the Deaf$^\dagger$                     & Newark, DE          & 5/30/13 \tabularnewline
Delaware School for the Deaf$^\dagger$                     & Newark, DE          & 5/31/13 \tabularnewline
National Technical Institute for the Deaf$^\ddagger$       & Rochester, NY       & 6/27/13 \tabularnewline
\bottomrule

\end{tabularx}

\caption{Dates and location information for AstroDance performances.  Collegiate performances are indicated by $^\ddagger$, secondary-school performances are indicated by $^\dagger$ while community performances are indicated by $^*$.}
\label{table:performances}

\end{table*}

\subsection{Demographic and open-ended responses}
\label{sec:survey_openended}

\begin{figure}[h!]
\begin{center}
\includegraphics[width=8cm,angle=0,clip=true]{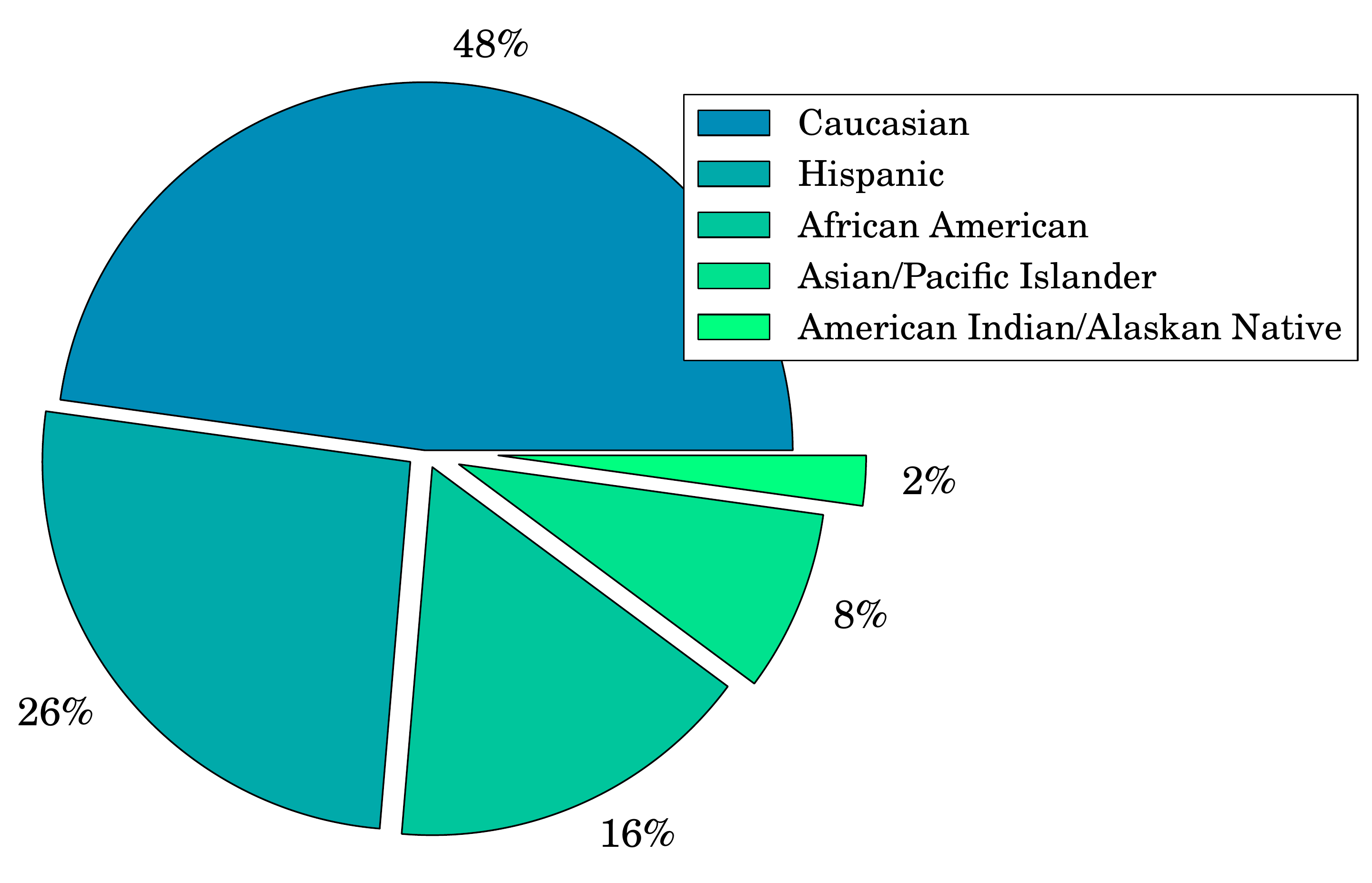}
\caption{Self-reported demographic information from surveyed audience members.
\label{fig:demographics}}
\end{center}
\end{figure}

Out of the 971 survey responses, 866 (89\%) self-reported demographic information on ethnicity.  Figure~\ref{fig:demographics} displays the ethnic breakdown with self-reporting of 44\% Caucasian, 26\% Hispanic, 16\% African American, 8\% Asian/Pacific Islander and 2\% Native Indian/Alaskan Native.  Self-reported ethnicity percentages in this survey are above the national averages for all minority (non-white) categories \citep{Vespa2018}.

The audience survey also included three free response questions.  Here, we summarize the results of each while including some specific illustrative response examples.  The first question, which generated 695 responses, was ``How would you describe the performance you saw to a friend or colleague?".  The bulk of the responses described a unique blending of art and science that positively impacted both aspects of the program.  Here we list three representative responses:

\begin{itemize}
\item {\it Different from regular performances I normally attend. There was narration, sign language interpretative audience interaction/participation, glow in the dark props. Yes, I learned that scientists and artists can work together to collaborate ideas/views.}

\item {\it A performance combining science and dance in an attempt to make astrophysics engaging to non-scientific folk.}

\item {\it Fantastic! Lovely and moving. Connecting science and dance is a wonderful idea.}
\end{itemize}

The second question asked ``Can you explain something new that you learned from the performance? If so, what did you learn?" and elicited 507 responses.  Topics mentioned included astrophysical objects such as black holes and supernovae, and how gravitational waves are generated and detected.  We provide three representative responses below:

\begin{itemize}
\item {\it A black hole drags everything with it as it rotates (like the sheet around the dancer).}

\item {\it I learned about the gravitational waves that are created from supernovae(sic) and converging black holes.}

\item {\it I learned the scientist measure gravity and matter through LIGO. I never heard of that before and will look it up!}
\end{itemize}

The final question sought additional comments on the performance and was presented as ``Any other comments about the performance?".  This question generated 377 responses of which we provide two below:

\begin{itemize}
\item {\it This is great, creative, beautiful and didactic --> do something please about cell biology.}

\item {\it Artistic expression is a great way to teach an understanding of complex. scientific concepts. Beautiful costume design \& props. Love the body movements forms!}
\end{itemize}

\section{Conclusions}
\label{sec:survey}

As part of the {\it AstroDance project}, after each performance, we collected data from the audience to statistically determine any differences in learning outcomes reported between the deaf/hard-of-hearing and hearing populations.  Our hypothesis was that since each performance incorporated various multimedia elements, including signed narration, the hearing- and DHH-audience members might report statistically similar results.

In our questionnaire, we surveyed demographic information including the audience member's age, gender, hearing status and race/ethnicity. Additionally, we included three five-point Likert scale questions and three open-ended questions designed to measure the audience's understanding of the physical concepts presented in the performance. We collected approximately 1000 responses, the majority of which were from students and split evenly between deaf/hard-of-hearing and hearing populations. Our responses were also split evenly between the male and female populations. What we found was enlightening. When asked: ``How much did you learn about science?", the DHH sample reported a statistically significant, higher-rating, suggesting that AstroDance enabled DHH audience members to successfully learn from the performance. Compared to most standard scientific presentations with no-signed interpretation, this is an exciting result as it demonstrates one method for actively engaging a DHH audience in scientific outreach.  In fact, a takeaway is that a multi-media experience that incorporates sign and visual elements benefits both DHH and hearing audiences and provides a roadmap for an inclusive and impactful science communication.

\section{Author's Note}
\label{sec:authorsnote}
Jason Nordhaus, Department of Science and Mathematics, National Technical Institute for the Deaf and Center for Computational Relativity, Rochester Institute of Technology; 52 Lomb Memorial Drive, Rochester, NY 14623; Manuela Campanelli, Center for Computational Relativity and Gravitation and Department of Mathematics, Rochester Institute of Technology; 52 Lomb Memorial Drive, Rochester, NY 14623, Email: manuela@astro.rit.edu; Joseph Bochner, Department of Cultural and Creative Studies, National Technical Institute for the Deaf, Rochester Institute of Technology; 52 Lomb Memorial Drive, Rochester, NY 14623; Thomas Warfield, Department of Performing Arts, National Technical Institute for the Deaf, Rochester Institute of Technology; 52 Lomb Memorial Drive, Rochester, NY 14623; Hans-Peter Bischof, Center for Computational Relativity and Gravitation and Department of Computer Science, Rochester Institute of Technology; 52 Lomb Memorial Drive, Rochester, NY 14623; Jake Noel-Storr, Undergraduate School of Science and Engineering and Kapteyn Astronomical Institute University of Groningen; Groningen, Netherlands.
.

Correspondence concerning this article should be addressed to Jason Nordhaus, National Technical Institute for the Deaf, 52 Lomb Memorial Drive, Rochester, NY 14623. Email: nordhaus@astro.rit.edu

\section{Acknowledgements}
\label{sec:acknowledgements}

This research was supported by the National Science Foundation Division of Research Learning award NSF DRL 1136221.  JN is supported by NSF award AST-1102738 and by NASA HST grant AR-12146.04-A.


\bibliography{References.bib}

\end{document}